\documentclass[aps,preprint,prd,showpacs,nofootinbib]{revtex4}

\usepackage{latexsym}
\usepackage{amsmath}
\usepackage{graphicx}
\usepackage{subfigure}
\usepackage{dcolumn}
\usepackage{bm}
\usepackage{amssymb}
\usepackage{color}
\usepackage{float}
\usepackage[colorlinks,linkcolor=magenta,anchorcolor=cyan,citecolor=blue]{hyperref}
\usepackage{appendix}

\def\be{\begin{equation}}
\def\ee{\end{equation}}
\def\ba{\begin{aligned}}
\def\ea{\end{aligned}}

\preprint{USTC-ICTS/PCFT-25-04}

\begin{document}

\title{Phase transition in a doubly holographic model of closed  $\mathbf{dS_{2} }$ spacetime }

\author{Wen-Hao Jiang$^{1,2}$\footnote{jiangwenhao16@mails.ucas.ac.cn}}
\author{Cheng Peng$^{3,6}$\footnote{pengcheng@ucas.ac.cn}}
\author{Yun-Song Piao$^{1,2,4,5}$\footnote{yspiao@ucas.ac.cn}}

\affiliation{$^1$ School of Physical Sciences, University of Chinese Academy
of Sciences, Beijing 100049, China}

\affiliation{$^2$ School of Fundamental Physics and Mathematical
    Sciences, Hangzhou Institute for Advanced Study, UCAS, Hangzhou
    310024, China}
    
\affiliation{$^3$ Kavli Institute for Theoretical Sciences (KITS), University of Chinese Academy of Sciences, Beijing 100190, China}

\affiliation{$^4$ International Center for Theoretical Physics
    Asia-Pacific, Beijing/Hangzhou, China}

\affiliation{$^5$ Institute of Theoretical Physics, Chinese
    Academy of Sciences, P.O. Box 2735, Beijing 100190, China}
\affiliation{$^6$ Peng Huanwu Center for Fundamental Theory, Hefei, Anhui 230026, China}

\begin{abstract}

Double holography has been proved to be a powerful method in comprehending 
the spacetime entanglement. In this paper we investigate the doubly holographic 
construction in ${\mathrm{dS_{2}} }$ spacetime. 
We find that in this model there exists a new extremal surface besides the 
Hartman-Maldacena surface and the island surface, which could lead to a more complex phase structure. We then propose a generalized mutual entropy to interpret the phase transition. However, this extremal surface has a subtle property that the length of a part of the geodesic is negative when this saddle is dominant. This is because the negative part of the geodesic is within the horizon of the bulk geometry. 
We move to the ${\mathrm{AdS_{2}} }$ spacetime and find this subtlety still exists. We purpose a simple solution to this issue.

\end{abstract}

\maketitle

\section{Introduction}\label{1}

   Black hole information paradox \cite{Hawking:1976ra} has been an important topic in theoretical physics. Recent studies in this field have helped us gain a deeper understanding about the entanglement and the emergence of spacetime \cite{Bousso:2022ntt}. One of the most crucial achievement is the reproduction of Page curve\cite{Page:1993wv,Page:2013dx} by the generalized entropy of radiation $S_R$ using the holographic island formula\cite{Penington:2019npb,Penington:2019kki,Almheiri:2019psf,Almheiri:2019hni,Almheiri:2019qdq,Almheiri:2019psy,Almheiri:2019yqk} for quantum extremal surface (QES) \cite{Engelhardt:2014gca}

   \begin{equation}\label{eq1.1}
    \mathit{S}_{\mathit{R}}=\mathrm{min} \{\mathrm{ext}[\frac{A(\partial I)}{4G_N}+S_{\mathrm{semi-cl} }(\mathrm{Rad} \cup I)]\},
   \end{equation}
   where $I$ is a region called island. This formula can be regarded as a generalized version of holographic entropy based on AdS/CFT correspondence. \cite{Ryu:2006bv,Ryu:2006ef,Hubeny:2007xt,Faulkner:2013ana}
\par
The island formula has been studied in various spacetime, and most of 
them involves islands in AdS spacetime 
\cite{Li:2021dmf,Gautason:2020tmk,Dong:2020uxp,Alishahiha:2020qza,Ling:2020laa,Matsuo:2020ypv,He:2021mst,Miao:2022mdx,Li:2023fly,Chang:2023gkt,Ahn:2021chg,Jeong:2023lkc,Liu:2022pan}. 
Since our universe is not in AdS spacetime \footnote{However, recently it has been found 
that the AdS spacetime might have imprinted unexpected signals in cosmologies,
e.g.\cite{Ye:2020btb,Jiang:2021bab,Ye:2021iwa,Wang:2024dka,Wang:2024hwd,Huang:2023chx,Cai:2023uhc}.}, there are several researches aiming at generalizing the island formula to de sitter spacetime and cosmology, see eg. recent \cite{Hartman:2020khs,Chen:2020tes,Balasubramanian:2020xqf,Aguilar-Gutierrez:2021bns,Levine:2022wos,Piao:2023vgm,Yadav:2022jib,Espindola:2022fqb,Ben-Dayan:2022nmb,Kames-King:2021etp,Aalsma:2021bit,Baek:2022ozg,Aalsma:2022swk,Teresi:2021qff,Seo:2022ezk,Azarnia:2021uch,Choudhury:2020hil,Choudhury:2022mch,Aguilar-Gutierrez:2023zoi,Aguilar-Gutierrez:2023ymx,Franken:2023ugu,Aguilar-Gutierrez:2023tic,Jiang:2024xnd}. However, there exist some subtleties when it comes to dS spacetime. For example, it is natural to introduce a cut-off surface in AdS so that for an observer outside the surface we can safely neglect the influence of gravity. For an enternal black hole in two dimensional $\mathrm{AdS_2}$ , we can collect the radiation in the asymptotic flat spacetime by placing the observer at a Minkowski bath and gluing it to the boundary of AdS spacetime. In contrast, in dS spacetime, we cannot easily find a place where gravity effect can be neglected. This is one of the problems that researchers are faced when implying the island formula in dS spacetime.
\par

\begin{figure}[H]
    \centering
    \includegraphics{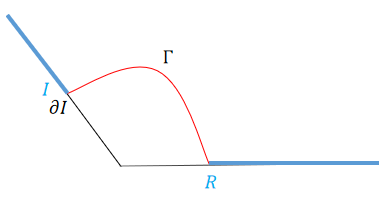}
    \caption{A schematic of (\ref{eq1.2}). Here $\Gamma$ is the extremal surface locating at the bulk that connect the end point of island region and radiation region, while the island is locating on the planck brane and $R$ is locating on the bath region, which is shown in blue in this figure.
    }
    \label{fig.1}
\end{figure}

    Recently another perspective of understanding the island formula called double holography \cite{Almheiri:2019hni,Suzuki:2022xwv,Chen:2020uac} emerges, which can be regarded as a combination of boundary conformal field theory (BCFT) \cite{Takayanagi:2011zk,Fujita:2011fp,Karch:2000gx} and brane world holography \cite{Randall:1999ee,Randall:1999vf,Gubser:1999vj,Karch:2000ct}. In this scenario a d-dimension gravity in AdS spacetime with its CFT couple to a bath is dual to a bulk (d+1)-dimension gravity which contains a dynamic boundary on a d-dimension planck brane \cite{Almheiri:2019hni}. Then in $d=2$ case we can reinterpret
(\ref{eq1.1}) as follows:
\begin{equation}\label{eq1.2}
    S_R=\mathrm{min}\{\mathrm{ext}[\frac{A(\partial I)}{4{G_N}^{(2)}}+\frac{\Gamma}{4{G_N}^{(3)}}
    ]\},
\end{equation}
where $\Gamma$ is the extremal surface that connects the island and the radiation region
(fig.\ref{fig.1}). This scenario may give us a

\par
    It is worth noting that this scenario has been verified in asymptotic AdS spacetime \cite{Fujita:2011fp,Suzuki:2022xwv,Takayanagi:2011zk,Izumi:2022opi,Geng:2022slq,Geng:2022tfc,Deng:2022yll,Liu:2023ggg}, but whether or how it could be applied to dS spacetime is still under discussion.
\par

    In this paper, our work is mainly based on the work \cite{Chang:2023gkt}. This model can be regarded as a generalized version of the doubly holographic framework which we introduce above by making the thermal bath on the flat brane embedded to the bulk spacetime. In this case, the gravity region and the bath are on a more equal status in the doubly holographic framework. When applying this model to the two dimensional closed $\mathrm{dS_2}$ spacetime, we can find that it is very similar to the situation of AdS spacetime. Then we can collect the radiation on the thermal bath where we can neglect the gravity effect. We discover that in this  scenario there exists another extremal surface besides the that connects the two baths and the island surface, and it is coincident with the Hartman-Maldacena (HM) surface \cite{Hartman:2013qma} that is mentioned in \cite{Shaghoulian:2021cef}. Then we propose a generalized mutual information to help understand the transition between phases where different surface has minimum area. However, our further analysis shows that a part of the geodesic of this saddle has a negative length as it locates in the horizon of the bulk geometry. This may be puzzling as we could introduce a bunch of negative geodesics and that may cause the entropy cannot be bounded from below. To solve the issue we put forward a proposal that only one of these geodesics with the same endpoints that contributes to the entropy.  Similar issue may also occur when we apply this model to $\mathrm{AdS_2}$ brane.

    Our paper is organized as follows: In Section.\ref{2} we review the process of embedding the $\mathrm{dS_2}$ spacetime in the \cite{Chang:2023gkt} and calculate the entanglement entropy of each extremal surface. In section.\ref{3}, we interpret the phase transition process by introducing a general mutual information and find the new saddle may dominate when the system is in low temperature. In Section.\ref{4} we discuss the subtlety of the new saddle and the origin behind this phenomenon. In Section.\ref{7}, we further discuss the implication of the subtlety of the saddle and make a proposal to fix the issue. In Section.\ref{6}, we move back to the spacetime of $\mathrm{AdS_2} $ eternal black hole and discuss the new saddle and the phase transition process in this situation. In Section.\ref{5}, we summarize our result and discuss the outlook of further research.

\section{Embedding the dS$_2$ brane in the AdS$_3$ bulk spacetime}\label{2}
\begin{figure}[H]
    \centering
    \includegraphics{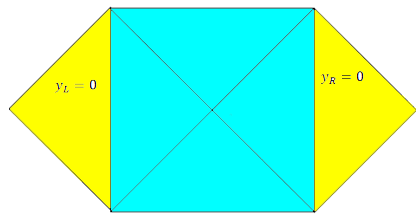}
    \caption{The Penrose diagram of the flat spacetime with thermal bath glued to the $dS_2$ spacetime. The $\mathrm{dS_2}$ spacetime is painted in cyan while the thermal bath is painted in yellow.
    }
    \label{fig.2}
\end{figure}

    In this section we follow \cite{Chang:2023gkt} to embed the spacetime into an $\mathrm{AdS_3}$ bulk and using formula (\ref{eq1.2}) to compute the entanglement entropy corresponding to each extremal surface. We consider Jackiw-Teitelboim (JT) gravity \cite{Jackiw:1984je,Teitelboim:1983ux} with positive cosmological constant with the following action:
\begin{equation}\label{eq2.1}
        S=\frac{\phi_{0}}{4\pi}\int_{\cal M} \sqrt{-g}R
        +\frac{1}{4\pi} \int_{\cal M}\sqrt{-g}\phi(R-2H)+S_{\partial {\cal M}},
\end{equation}
    where $H=\frac{2\pi}{\beta}=2\pi T$. The metric of the static patch of $\mathrm{dS_2}$ can be written in the form of:
\begin{equation}\label{eq2.2}
    ds^2=-(1-H^2 {\rho}^2)dt^2+\frac{d{\rho}^2}{1-H^2 {\rho}^2}
    =-H^2 \frac{dy^{+}dy^{-}}{(\cosh(\frac{H}{2}\left|y^{+}-y^{-}\right|))^2},
\end{equation}
    where $y^{\pm}=t \pm y$, $y=\pm \frac{1}{H}\tanh^{-1}(H\rho)$, and $\phi=H\phi_{r}\tanh(Hy)$.
As it is a $\mathrm{dS_2}$ spacetime, we can glue the $\mathrm{dS_2}$ brane and the flat bath along the $\rho =0$ line (Fig.\ref{fig.2}).

We rewrite the metric of $\mathrm{dS_2}$ spacetime and the flat bath into the conformal form by coordinate change $\omega^{\pm}=\pm \exp (\pm H y^{\pm}_{L}) , \omega^{\pm}=\mp \exp (\mp H y^{\pm}_{R})$, for the right side of the spacetime, the metric has the form:

\begin{equation}\label{eq2.3}
    \begin{aligned}
    ds^2
    &=-\frac{4}{1-\omega^{+}\omega^{-}}d\omega^{+}d\omega^{-}  \quad (y_R < 0),\\
    ds^2
    &=\frac{1}{H^2 \omega^{+}\omega^{-}}d\omega^{+}d\omega^{-} \qquad   (y_R > 0).
    \end{aligned}
\end{equation}
    When we embed the brane into a higher dimensional bulk, we should make sure that the metric of the brane and the induced metric of the bulk on the brane should satisfy:
\begin{equation}\label{eq2.4}
    g^{(d+1)}_{\mu\upsilon}|_{bdy}=\frac{1}{\epsilon^2}g^{(d)}_{\mu\upsilon}.
\end{equation}

    As the metric of bulk $\mathrm{AdS_3}$ spacetime can be written in $\mathrm{Poincar\acute{e}}$ form
    $ds^2=\frac{dz^2-d\omega^{+}d\omega^{-}}{z^2}$, we substitute this and (\ref{eq2.3}) into (\ref{eq2.4}), we can get:
\begin{equation}\label{eq2.5}
    \begin{aligned}
    z_{\mathrm{dS_2}}
    & =\frac{\epsilon_g}{2}|1-\omega^{+}\omega^{-}|, \\
    z_{bath}
    & = \epsilon_{b} H \sqrt{\omega^{+}\omega^{-}},
    \end{aligned}
\end{equation}
  this is how the  $\mathrm{dS_2}$ brane and the thermal bath embedded into the  $\mathrm{AdS_3}$ bulk. For the sake of convenience we set $\epsilon_{b} H=\epsilon_{g} = \epsilon $ for the rest of this paper.

\begin{figure}[H]
		\centering
		\includegraphics{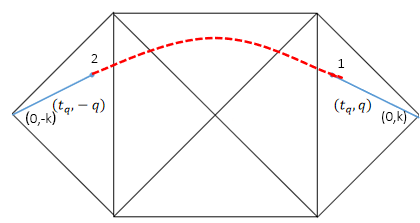}
		\caption{The schematic of the surface that connects the two sides of the system (red dashed line). Pay attention that the geodesic is in the bulk rather than on the $\mathrm{dS_2}$ brane.
		}
		\label{fig.3}
\end{figure}

	Before we dive into the entropy of each saddle we first calculate the entanglement entropy, $S_0$ of each radiation region, which will be helpful for the discussion in the next section. As shown in Fig.\ref{fig.3} , we set the 'inner' end points of radiation region, which are near the boundary of the dS spacetime at $(t_q,\pm q)$, and the 'outer' end points at $(0,\pm k)$ with $k \to \infty $. The entropy of each radiation region $S_0$ is
\begin{equation}\label{eq4.2}
		S_0=\frac{c}{6}\log[\frac{1}{\epsilon^2} \frac{(\omega^{+}_{q}-\omega^{+}_{k})(\omega^{-}_{q}-\omega^{-}_{k})}{\Omega(\omega_q)\Omega(\omega_k)}]
		=\frac{c}{6}\log[\frac{2}{\epsilon^2}(\cosh(Ht)-\cosh(Hq))].
\end{equation}
	As the radiation regions are locating in the flat spacetime, the entanglement entropy of a region is also the generalized entropy of this region.
	
    Then we calculate the  holographic entanglement entropy corresponding to different extremal surface. The first possible extremal surface is the surface that connect the two sides of the system (Fig.\ref{fig.3}). For the right side of the system, we substitute $(t_q,q)$ into (\ref{eq2.5}) and get:
\begin{equation}\label{eq2.6}
    z_{bath}=\epsilon \exp(Hq).
\end{equation}

    We label the near-boundary 'inner' end points of the radiation regions with '1' and '2'. Then we set:
\begin{equation}\label{eq2.7}
    (\omega_{i})_{t} = \frac{\omega^{+}_{i}+\omega^{-}_{i}}{2}
    (\omega_{i})_{x} =(-1)^{i}\frac{\omega^{-}_{i}-\omega^{+}_{i}}{2},
\end{equation}
    where $i=1,2 $. We can easily find that $(\omega_{1})_{t}=(\omega_{2})_{t}$, which makes the calculation of the length of the geodesic easier because the geodesic in the bulk $\mathrm{AdS_3}$ is a semi-circle which satisfies:
 \begin{equation}\label{eq2.8}
    z^2+\omega^2_x = z^2_{\ast}.
 \end{equation}
    We set $z=z_{\ast}\sin \theta, \omega_{x}=z\cos \theta$, then the geodesic length is:
 \begin{equation}\label{eq2.9}
    \Gamma= 2\int^{\frac{\pi}{2}}_{\frac{z_{\ast}}{z}}\frac{1}{\sin \theta} d\theta
    =2\log(\frac{2z_{\ast}}{z_{bath}})
    \approx 2\log(\frac{2\cosh (Ht)}{\epsilon}).
\end{equation}
    Substituting this into (\ref{eq1.2}), the holographic entanglement entropy corresponding to this geodesic is:
\begin{equation}\label{eq2.10}
    S_1=\frac{c}{3}\log(\frac{2\cosh (Ht)}{\epsilon}).
\end{equation}
It grows monotonically with time.

\begin{figure}[H]
    \centering
    \includegraphics{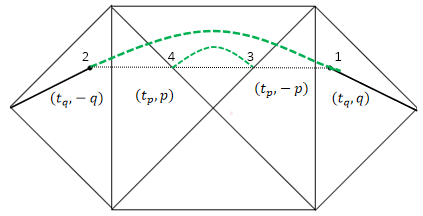}
    \caption{The schematic of the surface that connects the endpoints '1' with '2' and '3' with '4' (green dashed lines). Again the geodesic is in the bulk rather than on the $\mathrm{dS_2}$ brane. Noticing that in this situation the two sides of the system is still connected.
    }
    \label{fig.4}
\end{figure}

    There is another possible extremal surface whose endpoints are on the $\mathrm{dS_2}$ brane, we label the right and left endpoints on  $\mathrm{dS_2}$ brane '3' and '4'. Their coordinate are $(t_p, -p)$ and $(t_p, p)$. We first consider the situation where the geodesics connect '1' with '2' and '3' with '4' (Fig.\ref{fig.4}). The calculation process is merely the same with the first situation. The entanglement entropy in this situation is:
\begin{equation}\label{eq2.11}
    S_2=\frac{2c}{3}\log(\frac{2\cosh (Ht)}{\epsilon})+\frac{c}{3}\log{\frac{1}{\cosh(Hp)}}+2H\phi_{r}\tanh(Hp).
\end{equation}

 Since the quantum extremal surface is the maximin surface \cite{Hubeny:2007xt}, we should maximize the surface in time and minimize it in space. To make this entanglement entropy minimum in space direction, we need to make $p \to \infty$. As $p$ is in $y$ direction, when $p \to \infty$, $\rho \to \frac{1}{H}$. That means in this situation the endpoints on the $\mathrm{dS_2}$ brane '3' and '4' are just on the horizon of static patchs, which seems coincident with the HM surface of dS spacetime in \cite{Shaghoulian:2021cef}.

    To make $S_2 $ finite, we introduce another cut-off '$ \varepsilon $' as we take the limit $p \to \infty$. Now the formula of $S_2$ becomes:
\begin{equation}\label{eq2.12}
    S_2=\frac{2c}{3}\log(\frac{\varepsilon\cosh (Ht)}{\epsilon})+2H\phi_{r}.
\end{equation}

\begin{figure}[H]
    \centering
    \includegraphics{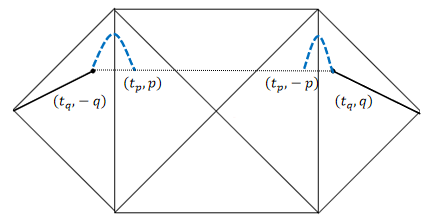}
    \caption{The schematic of the island surface that connects the endpoints '1' with '3' and '2' with '4' (blue dashed lines). The geodesic is still in the bulk rather than on the $\mathrm{dS_2}$ brane. It seems that in this situation the two sides of the system are no longer connected.
    }
    \label{fig.5}
\end{figure}

    Next we consider the geodesics that connect '1' with ' 3' and '2' with '4' (Fig.\ref{fig.5}). As $(\omega_{1})_{t} \neq \omega_{3})_{t}$, the length of geodesic in this situation is:
\begin{equation}\label{eq2.13}
    \frac{\Gamma}{4G_N} = \frac{c}{6}\log(\frac{(\omega^{+}_{1}-\omega^{+}_{3})(\omega^{-}_{3}-\omega^{-}_{1})}{z_{bath}z_{\mathrm{dS_2}}})=
    \frac{c}{6}\log(\frac{2[\cosh[H(p+q)]-\cosh[H(t_p-t_q)]]}{\epsilon^2 \cosh(Hp)}).
\end{equation}
    To make the entanglement entropy maximum in time and minimum in space, we should take $t_p=t_q$ and $p=q$. Then we have:
\begin{equation}\label{eq2.14}
    S_3=\frac{2c}{3}\log(\frac{2}{\epsilon}\sinh(\frac{Hq}{2})).
\end{equation}
    This is the well known island surface, with the entanglement entropy saturates in time.

\section{Phase transition and its interpretation by generalized mutual information}\label{3}

    After calculating the entanglement entropy corresponding to each kind of extremal surfaces, we are now able to analyze the phases and the transition among them in this system. There are three different possible extremal surfaces, and the dominant surface will be the one that leads to the minimum general entanglement entropy, which means:
 \begin{equation}\label{eq3.1}
    S=\min[S_1,S_2,S_3].
 \end{equation}
 
\begin{figure}[H]
 	\centering
 	\includegraphics{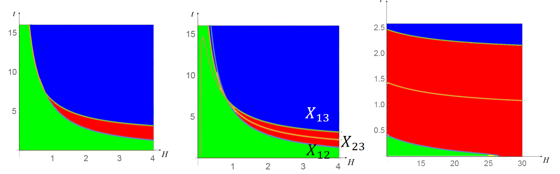}
 	\caption{$\bold{Left}$: The phase transition diagram with the parameter set by $\varepsilon = \epsilon =0.01, d=\frac{3\phi_{r}}{c}= 0.1 , q = 0.2$ \protect\footnotemark. The phase which $ S_1 $ dominates is painted in red, while the $ S_2 $ dominant phase is painted in green, and $ S_3 $ dominant phase is painted in blue. It can be seen that there exists a critical temperature, where three dominant phase join together. If the temperature of the system is lower than this critical point, $S_1 $ saddle will never dominate. $\bold{Middle} $: The phase transition diagram painted with critical points between each pairs of different extremal surfaces. $\bold{Right}$: There exists another critical temperature is from which $S_2 $ will never dominate, meaning that there only exists transition between $S_1 $ and $S_3 $ at high temperature limit. This is coherent with the result in \cite{Chang:2023gkt}.
 	}
 	\label{fig.6}
 \end{figure}

In Fig.(\ref{fig.6}), we plot the region where $S_i$ dominates.
 
 \footnotetext{Here we may set $q$ and the cut-off a little bit large in order to have a sharper figure. In this paper what we intend to discuss is the situation that q is very small}
 
   Comparing formula (\ref{eq2.10}),(\ref{eq2.12}) and (\ref{eq2.14}) we get the expression of critical points $X_{ij}$ between each pair of extremal surfaces, where $i$ and $j$ are defined by $S_i $, $S_j $ :
 \begin{equation}\label{eq3.2}
    \begin{aligned}
        X_{12}
        &:\cosh(Ht)=\frac{2\epsilon}{\varepsilon^2}\exp[-\frac{6H\phi_{r}}{c}],\\
        X_{23}
        &:\cosh(Ht)=\frac{2}{\varepsilon}\sinh[\frac{Hq}{2}]\exp[-\frac{3H\phi_{r}}{c}],\\
        X_{13}
        &:\cosh(Ht)=\frac{2}{\epsilon}(\sinh[\frac{Hq}{2}])^2.
    \end{aligned}
\end{equation}
	These critical points form curves in the $H-t $ diagram and with the help of them we can plot the middle diagram of Fig.\ref{fig.6}. 
    
   Another way to understand the phase transitions from the connected phases to the island phase is in terms of mutual information \cite{Geng:2020kxh,RoyChowdhury:2023eol,Saha:2021ohr,RoyChowdhury:2022awr,Kudler-Flam:2021alo,Kudler-Flam:2022zgm}. For two subsystems $A$ and $B$, the mutual information between them is:
\begin{equation}\label{eq4.1}
    I(A;B)=S(A)+S(B)-S(AB).
\end{equation}
  The relation between the mutual information and the Page curve was previously studied in \cite{RoyChowdhury:2022awr}. Since the Page transition can be regarded as a phase transition between two saddles, it is therefore natural to use the mutual information to interpret the phase transitions related to the island phase in our setup. We identify the radiation regions in the two sides of the systems as the two subsystems $A$ and $B$, then we propose a generalized mutual information with the form:
\begin{equation}\label{eq4.9}
	\begin{aligned}
		I_{gen}(A;B)
		&:=S_{gen}(A)+S_{gen}(B)-S_{gen}(AB) \\
		&=2S_0-S_i (i=1,2) \\
		&\equiv I_i \quad . \\ 
	\end{aligned}
\end{equation}
 It is worth noticing that when $i=1 $, $I_i $ coincide with $I(A;B) $.
 
First let's assume that the $S_2$ saddle does not exist, then we mainly focus on $X_{13} $ and the phase transition process will be like:

\begin{figure}[H]
    \centering
    \includegraphics{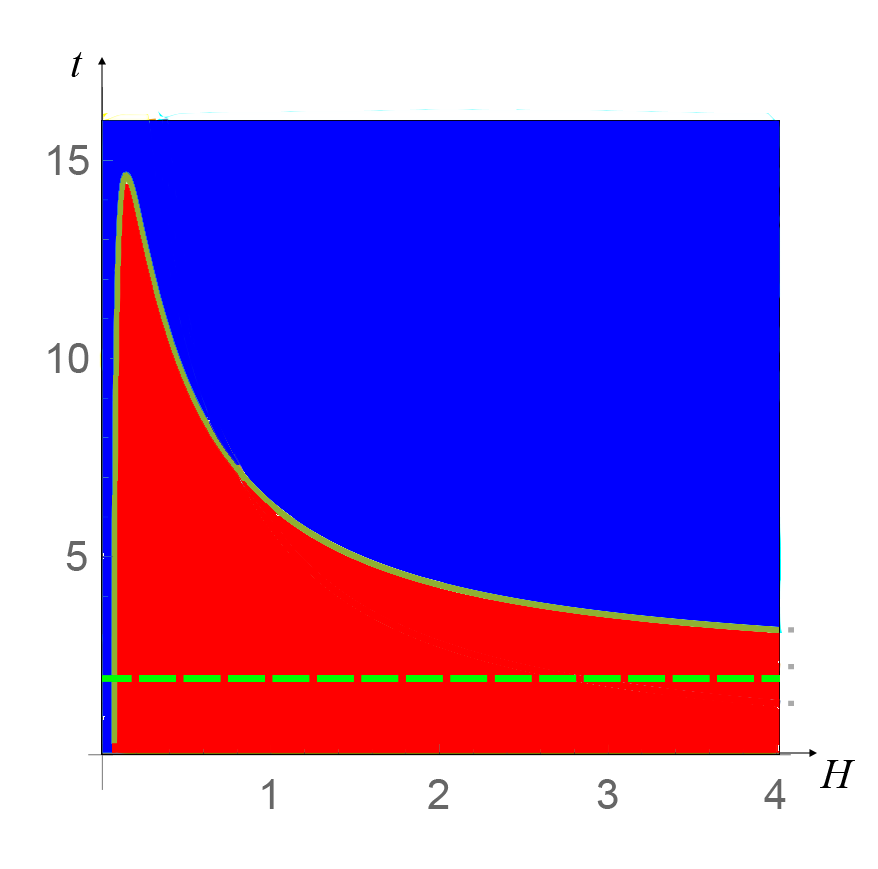}
    \caption{Phase diagram of the system if we remove $S_2$. It is obvious that the boundary line between the $S_1$ dominant area and $S_3$ dominant area ($X_{13} $) deviates from the asymptotic line (the green dashed line) when it is at the low temperature region. }
    \label{fig.10}
\end{figure}

As suggested in \cite{RoyChowdhury:2022awr} the transition between the connected saddle and the island saddle happens when $I(A;B)=0 $. In our setup, the analogous transition is between the $S_1$ and the $S_3$ saddles. 

\begin{figure}[H]
    \centering
    \includegraphics[width=8cm]{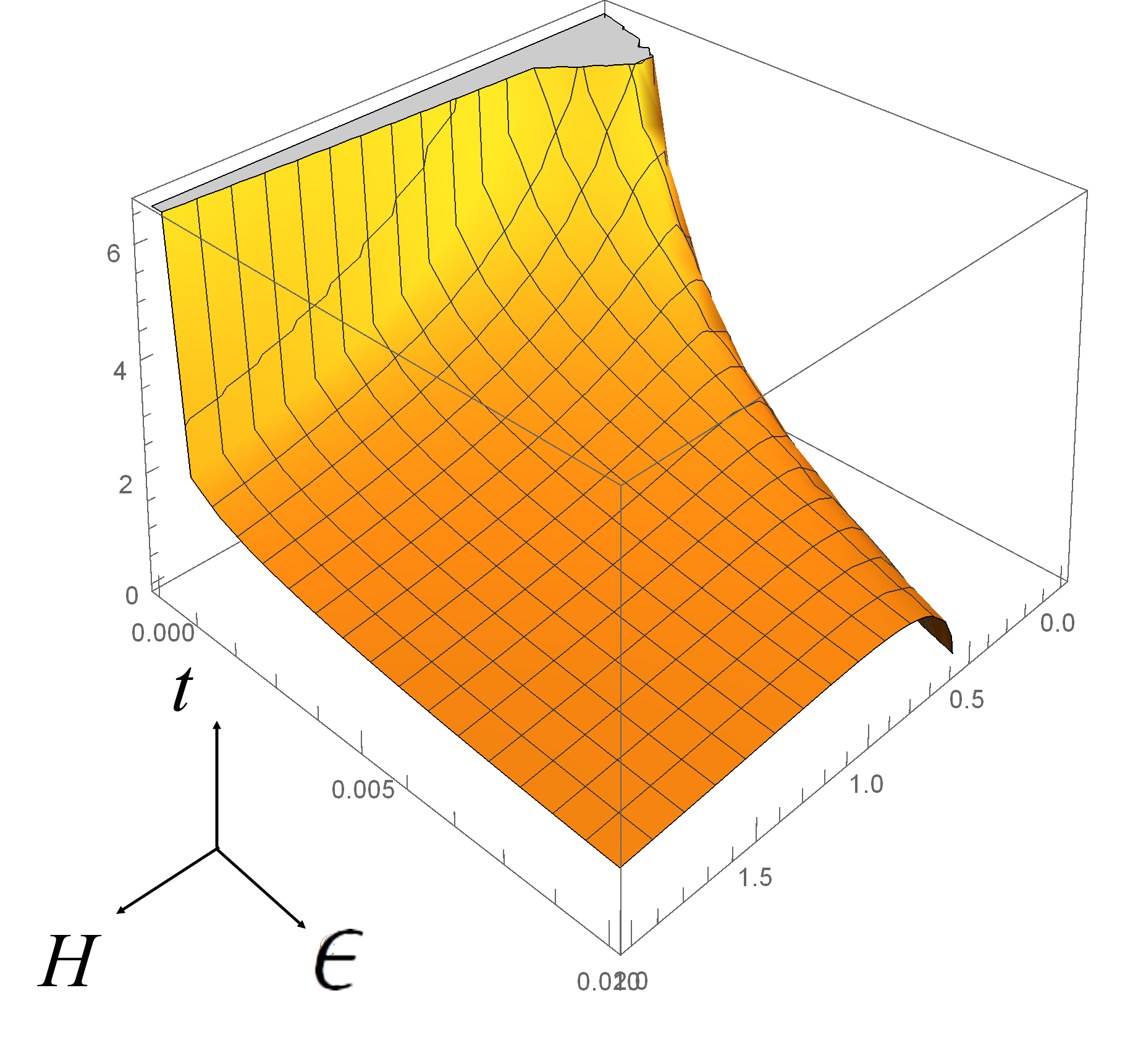}
    \caption{The curve of $X_{13}$ in different value of the cut-off. It can be found obviously that as the cut-off $ \epsilon $ gets smaller, the divergence of $X_{13}$  from the line that satisfies $I_1=0$ grows larger. }
    \label{fig.11}
\end{figure}

We can compare these transition points with the generalized mutual information. When  $S_1$ dominates we have:
\begin{equation}\label{eq4.3}
    I_1
    =2S_0-S_1
    =\frac{c}{3}\log[\frac{(\cosh(Ht)-\cosh(Hq))}{\epsilon\cosh(Ht)}].
\end{equation}
$ I_1 =0 $ happens at:
\begin{equation}\label{eq4.4}
    t=\frac{1}{H}\cosh^{-1}(\frac{\cosh(Hq)}{1-\epsilon}),
\end{equation}
which means when $I_1=0$, $t \approx q$. From (\ref{eq3.2}) we can tell that it is also the asymptotic line of the boundary between $S_1$ and $S_3$ ($X_{13}$ ) at high temperature limit. Therefore, at high temperature region, when the generalized mutual information $ I_1 $ becomes zero, the transition between $S_1$ and $S_3$ occurs. However, in the low temperature region $X_{13}$  deviates from $t=q$ sharply. \footnote{It is not the cut-off effect. In contrast, the smaller the cut-off is, the larger the deviation become (Fig.\ref{fig.11})}. One possible interpretation of this deviation at low temperature is the existence of another saddle: $S_2$.

Now we take $S_2$ saddle into account. We will show how to interpret the phase transition between $S_2$ and $S_3$ in the low temperature region through the generalized mutual information. When $S_2 $ dominates, the generalized mutual information in this situation is:
\begin{equation}\label{eq4.5}
    I_2
    =2S_0-S_2
    =\frac{c}{3}\log[\frac{2(\cosh(Ht)-\cosh(Hq))}{\epsilon^2(\cosh(Ht))^2}]-2H\phi_{r}.
\end{equation}
Rewrite the form of $X_{23} $ in (\ref{eq3.2}) and compare it with $I_2=0$ :
\begin{equation}\label{eq4.6}
    \begin{aligned}
        2[\cosh(Ht)-\cosh(Hq)]
        & =\epsilon^2(\cosh(Ht))^2\exp(2Hd), \\
        2\sinh(\frac{Hq}{2})
        & =\epsilon\cosh(Ht)\exp(Hd)
    \end{aligned}.
\end{equation}
We find when $X_{23}$ and $I_2=0$ coincide, it should satisfy:
\begin{equation}\label{eq4.7}
    \cosh{Ht}-\cosh(Hq)=2(\sinh(\frac{Hq}{2}))^2
    =\cosh(Hq)-1 \approx 0,
\end{equation}
which is approximately correct when both of $H$ and $q$ are small. It means that the condition $I_2=0$ aligns with the phase transition points between $S_2$ and $S_3$ occurs at low temperature, indicating the generalized mutual information is helpful in the interpretation of the phase transition process.

\section{$\mathbf{S_{2} }$ saddle revisit: island inside the horizon}\label{4}

From our analysis in Section.\ref{3}. The $S_{2}  $ saddle is very likely to be dominant at early time when the system is in relatively low temperature, and the phase transition at in different temperature is shown in fig.\ref{fig.6}. However, the $S_2 $ saddle has a subtle property. $S_2 $ saddle consists of two geodesics. The length of this two geodesics can be written as:

\begin{equation}\label{eq7.1}
    \begin{aligned}
        L_{12}
        &= 2\log(\frac{2\cosh (Ht)}{\epsilon}) \\
        L_{34}
        &= 2\log(\frac{2\cosh (Ht)}{\epsilon \cosh(Hp)})
        = 2\log(\frac{2{\varepsilon}^{2}\cosh (Ht)}{\epsilon}).\\
    \end{aligned}
\end{equation}
From (\ref{eq2.10}) and (\ref{eq2.12}) we can know that $S_2 - S_1 = \frac{c}{3}\log(\frac{2{\varepsilon}^{2}\cosh (Ht)}{\epsilon})+2H\phi_{r} $. It requires $S_2 - S_1 < 0 $ when $S_2 $ saddle is dominant, however that will certainly lead to $L_{34} < 0 $. A geodesic with negative length may be pathological.
    To probe the origin of this issue, we notice that the bulk spacetime where the geodesics lie. From \cite{Chang:2023gkt}, the bulk spacetime is a BTZ black hole with metric:
\begin{equation}\label{eq7.2}
ds^2 = -(r-r_0)dt^2 + \frac{dr^2}{r-r_0} + r^2 dx^2,
\end{equation}
where $r_0 = H $. Then the position of the $\mathrm{dS_2} $ brane in the BTZ bulk will be:
\begin{equation}\label{eq7.3}
    r=\frac{H}{\epsilon \cosh (Hp)} = \frac{H{\varepsilon}^2}{\epsilon}.
\end{equation}
From (\ref{eq7.1}) and (\ref{eq7.3}) we can find when $L_{34} < 0 $, $r < r_0 $, meaning that point '3' and '4' are inside the horizon of the bulk BTZ black hole. That explains why the length of this part of geodesic is negative.

\section{Implication of the negative geodesic}\label{7}

    In Section.\ref{4} we have found that when $S_2 $ saddle is dominant, it requires the geodesic that connects the endpoints in $\mathrm{dS_2} $ brane locating inside the horizon in the bulk BTZ spacetime, which makes the geodesic have a negative length. Negative geodesics may cause this saddle  unstable, we may consider configuration with a bunch of geodesics with endpoints $(t,p_i),(i=1,2,...,n) $ inside the horizon in the bulk. In this situation, the holographic entropy of this saddle is:
\begin{equation}\label{eq8.1}
    S_{\Sigma} =\frac{(n+1)c}{3}\log(\frac{2\cosh (Ht)}{\epsilon})+\Sigma_{i=1}^{n} (\frac{c}{3}\log{\frac{1}{\cosh(Hp_i)}}+2H\phi_{r}\tanh(Hp_i)).
\end{equation}
    Extremalizing it with each $p_i $, we will find that when $p_i \to \infty $ the entropy becomes minimum, and that will make these geodesics coincide whose endpoints locate near the horizon of dS brane. From the result in Section.(\ref{4}), these geodesics have negative length. This means that when $n $ gets larger, $S_{\Sigma} $ will get smaller, which means that $S_{\Sigma} $ lacks a lower bound.

    To resolve this problem, we supplement the island formula with a further requirement that contributions from bulk geodesics should be counted without multiplicity, in other words, geodesics with the same limiting end points contribute only once to the entropy. With this condition, we get sensible results of the entropy. 

\section{Moving back to AdS$_2$ eternal black hole}\label{6}

    In previous sections we have discussed the new saddle and the phase transition of the model from \cite{Chang:2023gkt} in $\mathrm{dS_2} $ spacetime, and we analyze the phase transition from the perspective of the generalized mutual information. In this section we returned to the eternal black hole situation in $\mathrm{AdS_2} $ spacetime and explore whether saddles like $S_2$ and similar phase transitions still exist.

  \begin{figure}[H]
    \centering
    \includegraphics{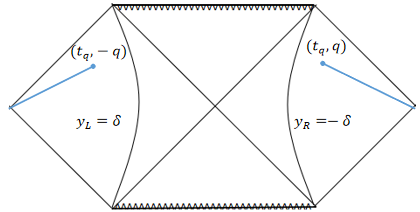}
    \caption{The Penrose diagram of the two sided eternal black hole in $\mathrm{AdS_2}$, for the right system $y_R=-\delta $ is the cut-off surface where the $\mathrm{AdS_2}$ brane and the heat bath are glued together}
    \label{fig.12}
\end{figure}
    We glue a flat spacetime as a heat bath along a cut-off surface $y_R=-\delta$ at the right side of the system (and $y_L=\delta$ on the left side ) (See Fig.\ref{fig.12}), then the metric of the eternal black hole is:
\begin{equation}\label{eq6.1}
    ds^2=-H^{2}\frac{\sinh^{2}(H\delta)}{\sinh^{2}(H\frac{y_{R}^{-}-y_{R}^{+}}{2})}dy_{R}^{+}dy_{R}^{-}.
\end{equation}
    Applying the coordinate transition $\omega^{\pm}=\pm \exp (\pm H y^{\pm}_{R}) , \omega^{\pm}=\mp \exp (\mp H y^{\pm}_{L})$, the metric of the right half of the spacetime is:
\begin{equation}\label{eq6.2}
    \begin{aligned}
    ds^2
    &=-\frac{4\sinh^{2}(H\delta)}{1+\omega^{+}\omega^{-}}d\omega^{+}d\omega^{-}  \quad (y_R < -\delta),\\
    ds^2
    &=\frac{1}{ \omega^{+}\omega^{-}}d\omega^{+}d\omega^{-} \qquad \qquad   (y_R > -\delta).
    \end{aligned}.
\end{equation}

    The calculation of holographic entanglement entropy is similar to those in Sec.\ref{2}, and the results are:
\begin{equation}\label{eq6.3}
    \begin{aligned}
        S_{1}^{'}
        &=\frac{c}{3}\log[\frac{2}{\epsilon}\cosh(Ht)],\\
        S_{2}^{'}
        &=\min_{p}[\frac{2c}{3}\log(\frac{2\sqrt{\sinh(H\delta)}\cosh(Ht)}{\epsilon})+\frac{c}{3}\log(\frac{1}{\sinh(Hp)})+2H\phi_{r}\frac{1}{\tanh(Hp)}],\\
        S_{3}^{'}
        &=\min_{p}[\frac{c}{3}\log[\frac{4\sinh^{2}(\frac{H(p+q)}{2})\sinh(H\delta)}{\epsilon^{2}\sinh(Hp)}]+2H\phi_{r}\frac{1}{\tanh(Hp)} ],
    \end{aligned}.
\end{equation}
here we have set $\epsilon_{b}=\epsilon_{g}=\epsilon $. It can be seen easily that $   S_{2}^{'} $ minimizes at  $p \to \infty $. So similar to what we have done in Sec.\ref{2}, we introduce another cut-off $\varepsilon$ and evaluate the $\min_p $ to rewrite $ S_{2}^{'} $ in the form of:
\begin{equation}\label{eq6.4}
    S_{2}^{'}=\frac{2c}{3}\log[\frac{\varepsilon\sqrt{\sinh(H\delta)}\cosh(Ht)}{\epsilon}]+2H\phi_{r}.
\end{equation}
    The $p $ dependent term in $ S_{2}^{'} $ correspond to a geodesic whose end points locate on the horizon of the eternal black hole in the $p \to \infty $ limit. Therefore, the saddle we mainly pay attention to still exists in $\mathrm{AdS_2} $ spacetime.

   The extremal of $S_{3}^{'} $ appears at $\frac{\partial{S_{3}^{'}}}{\partial{p}}=0 $, which leads to:
\begin{equation}\label{eq6.5}
    \frac{\sinh(\frac{H(p-q)}{2})}{\sinh(\frac{H(p+q)}{2})}=\frac{2Hd}{\sinh(Hp)},
\end{equation}
    where $d=\frac{3\phi_{r}}{c} $. Generally this equation is not easy to solve. To the limit $q \to 0$, the equation becomes $\sinh(Hp) = 2Hd $. Substituting it back to
(\ref{eq6.4}) we get:
\begin{equation}\label{eq6.6}
    S_{3}^{'}=\frac{c}{3}\log[\frac{\sqrt{4H^{2}d^{2}+1}-1}{\epsilon^{2}Hd}\sinh(H\delta)]+2H\phi_{r}\sqrt{\frac{1}{4H^{2}d^{2}}+1}.
\end{equation}
     Similar to the process in Sec.\ref{3}, we calculate the transition between each saddles, label as $X_{ij}^{'}$:
 \begin{equation}\label{eq6.7}
    \begin{aligned}
        X_{12}^{'}
        &:\cosh(Ht)=\frac{2\epsilon}{\varepsilon^{2}\sinh(H\delta)}\exp[-2Hd],\\
        X_{23}^{'}
        &:\cosh(Ht)=\frac{2}{\varepsilon}\sqrt{\frac{\sqrt{4H^{2}d^{2}+1}-1}{4Hd}} \exp[-Hd(1-\sqrt{\frac{1}{4H^{2}d^{2}}+1})],\\
        X_{13}^{'}
        &:\cosh(Ht)=\frac{2}{\epsilon}\frac{\sqrt{4H^{2}d^{2}+1}-1}{4Hd}\sinh(H\delta)\exp[\sqrt{1+4H^{2}d^{2}}].
    \end{aligned}
\end{equation}
   Then we get the phase diagram:

\begin{figure}[H]
    \centering
    \includegraphics[width=8cm]{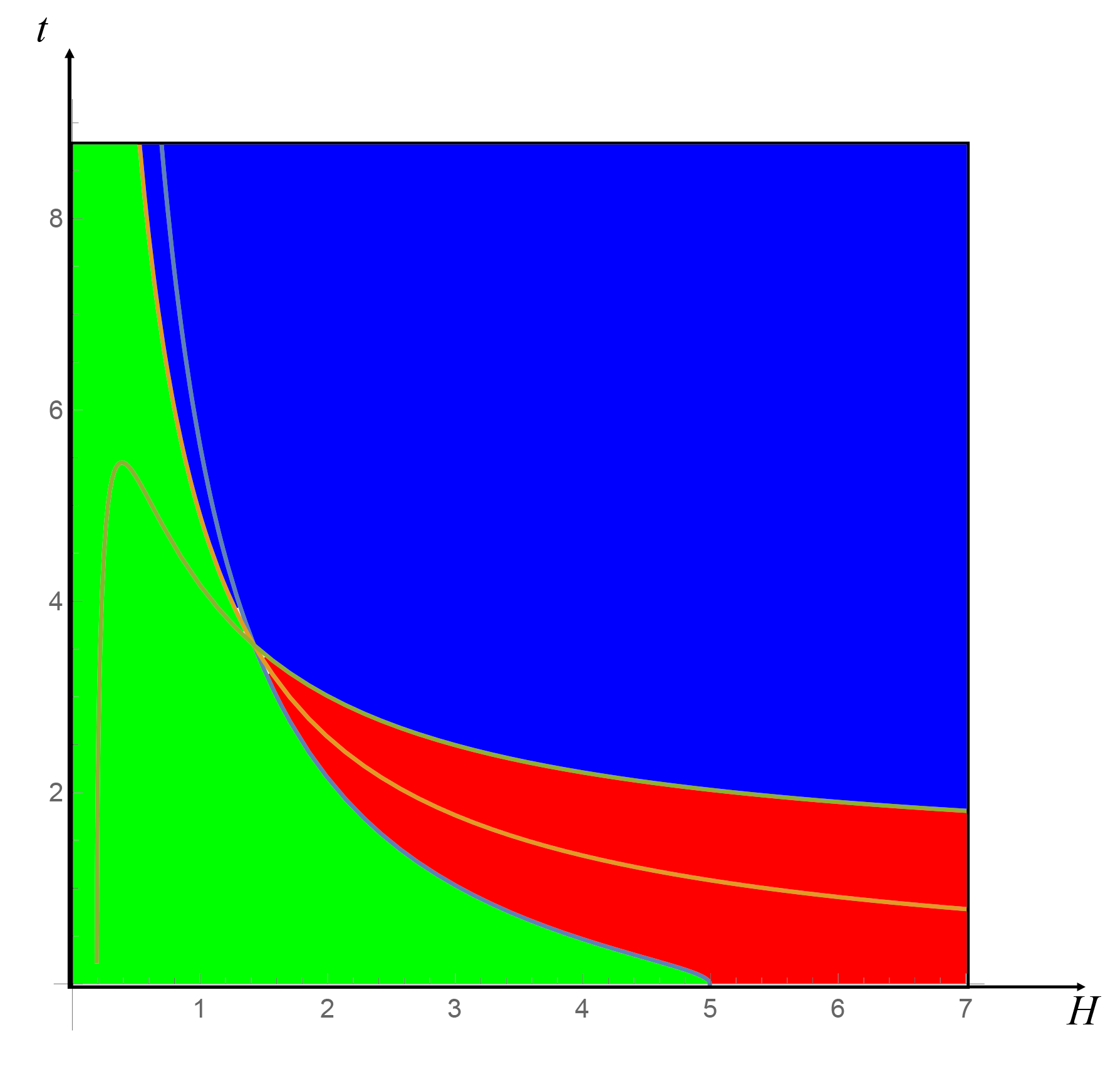}
    \caption{The phase transition diagram with the parameter set by $\varepsilon = \epsilon =0.01, d=\frac{3\phi_{r}}{c}= 0.1$. The phase which  $ S_{1}^{'} $ dominates is painted in red, while  $ S_{2}^{'} $dominant phase is painted in green, and  $ S_{3}^{'} $ dominant phase is painted in blue. It can be seen that there exists a critical temperature $H_{1}^{'}$ where three dominant phase join together. And another critical temperature $H_{2}^{'}$, from which $ S_{2}^{'} $ no longer dominates. 
    }
    \label{fig.13}
\end{figure}

    Compare Fig.\ref{fig.13} with Fig.\ref{fig.6} it is evident that the phase transition process of $\mathrm{AdS_2} $ eternal black hole in this model is similar to that in $\mathrm{dS_2} $ spacetime. The critical temperature where three phases joint together $H_{1}^{'}$ and the critical temperature from which $ S_{2}^{'} $ never dominates satisfy:
 \begin{equation}\label{eq6.8}
    \begin{aligned}
        H_{1}^{'}
        &:4H_{1}^{'}d\exp(-2H_{1}^{'}d)=(\sqrt{4H_{1}^{'2}d^{2}+1}-1)\sinh^{2}(H_{1}^{'}\delta)\exp(\sqrt{1+4H_{1}^{'2}d^{2}}),\\
        H_{2}^{'}
        &:\exp(2H_{2}^{'}d)\sinh(H_{2}^{'}\delta)=\frac{2}{\epsilon}.
    \end{aligned}
\end{equation}
    Now we compare the phase transition with the generalized mutual information. Similar to the process in Sec.\ref{4}, we assume that there exists another endpoint at the right bath region locating at $(0,\frac{1}{\xi})$ (for the left side is at $(0,-\frac{1}{\xi})$), where $\xi \to 0 $. Then the radiation entropy of one side can be written in the form of:
\begin{equation}\label{eq6.9}
    S_{0}^{'}
    =\frac{c}{6}\log[\frac{2}{\epsilon^2}|\cosh(Ht)-\cosh(\frac{H}{\xi})|].
\end{equation}
    Then the generalized mutual information when $ S_{1}^{'} $ is dominant would be:
\begin{equation}\label{eq6.10}
        I_{1}^{'}
        =2S_{0}^{'}-S_{1}^{'}
        =\frac{c}{3}\log[\frac{(\cosh(Ht)-\cosh(\frac{H}{\xi}))}{\epsilon\cosh(Ht)}].
\end{equation}
    Solving $t $ when $I_{1}^{'}=0 $ gives $t \approx \frac{1}{\xi} \to \infty $. Considering that the asymptotic line of $X_{13}^{'} $ at high temperature is $t= \delta + 2d \to \infty $, when $d $ is very large. There could be a finite difference between $t=\frac{1}{\xi} $ and $t= \delta + 2d $ that is not captured in our calculation.

    The problem we have studied in Sec.\ref{4} of  $S_{1}^{'} $ saddle at low temperature limit still occurs in this situation, which also indicates the existence of $S_{2}^{'} $ saddle at low temperature. As $I_{2}^{'}=2S_{0}^{'}-S_{2}^{'}=0 $
    requires:
\begin{equation}\label{eq6.14}
    \frac{2|\cosh(Ht)-\cosh(\frac{H}{\xi})|}{\epsilon^{2}\sinh(H\delta)\cosh^{2}(Ht)}=e^{2Hd}.
\end{equation}
	Solving $t $ in (\ref{eq6.14}) in the $H \to 0 $ limit we can get:
\begin{equation}\label{eq6.17}
	t \sim \sqrt{\frac{1}{\xi^2}+\frac{\epsilon^2}{H}} \to \infty.
\end{equation}	
	 It is worth noting that at low temperature limit the formula of $X_{23}^{'} $ can be approximated as:
\begin{equation}\label{eq6.15}
\cosh(Ht)=\frac{2}{\epsilon}\sqrt{\frac{Hd}{2}}e^{\frac{1}{2}}.
\end{equation}
    When the line $X_{23}^{'} $ and the line $I_{2}^{'}=0 $ coincide, meaning that the dominant saddle changes from $S_{2}^{'} $  to the island surface when the generalized mutual information reduces to zero, we can substitute (\ref{eq6.15}) back to (\ref{eq6.14}), then at low temperature we can get :
\begin{equation}\label{eq6.16}
	t \sim \sqrt{\frac{1}{\xi^2}+\delta d} \to \infty.
\end{equation}

    Comparing with (\ref{eq6.17}) we can find that when $H \to 0 $, for we focus on the low temperature region, $X_{23}^{'} $ matches $I_2^{'}=0 $ qualitatively in the sense that both (\ref{eq6.17}) and (\ref{eq6.16}) diverge. This indicates that the generalized mutual information is helpful in understanding the transition between  $S_{2}^{'} $ surface and island saddle at low temperature limit.

    In this section we find that the saddle like $S_2 $ in $\mathrm{dS_2} $ still exists in $\mathrm{AdS_2} $ spacetime, indicating that there should also be a similar phase transition process. We analyze the phase transition process in double sided $\mathrm{AdS_2} $ eternal black hole and interpret it with the method of generalized mutual information. In this situation the mutual informaton scenario may not fit exactly like in $\mathrm{dS_2} $ spacetime and need some approximations. However, it still indicates that the system may not only include $S_{1}^{'} $ saddle and $S_{3}^{'} $ saddle (island surface). The issue we have discussed in Section.\ref{7} still exists in this situation, as $S_{2}^{'} $ saddle also includes a geodesic with negative length when it is dominant. We could also use the proposal we have discussed in Section.\ref{7} to fix this issue.

\section{Discussion}\label{5}
    The island formula proves to be successful in deepening our understanding of information paradox in Anti de Sitter spacetime, and there have been generalization of it to a vast range of spacetime, in order to get the answers closer to the reality. When it comes to de Sitter spacetime, gravity cannot be ignored everywhere, making it subtle to place the observer to collect the radiation. Although the model of \cite{Chang:2023gkt} has strict constraints when applying to dS spacetime, it has  advantages that we can safely put the observer in the flat bath where the gravitational effect can be ignored. In this paper, we take a step further and analyze the phase transition of the $\mathrm{dS_2}$ spacetime in this model, and find that there may exist another kind of extremal surface in this model. Considering that this saddle  still connects the two sides of the $\mathrm{dS_2}$ spacetime, the transition from this saddle to the island surface is in agreement with the transition that is mentioned in \cite{Hartman:2013qma}. In addition, this saddle is partly coincide with the argument about the HM surface in dS spacetime in \cite{Shaghoulian:2021cef}, since this saddle contains a geodesic that links the same piece of the two sides of the dS spacetime. We wonder whether it could be regarded as a kind of HM surface.

    After analyzing the entropy of different saddles. We use the method of generalized mutual information to interpret the phase transition. For a two sided system at the early time the dominant saddle is the connected one while at the late time it will switch to disconnect. This is a phase transtion process that is mentioned in \cite{Hartman:2013qma}. It seems that mutual information could be inspiring in studying this kind of issue. When we temporarily remove the new saddle that we are interested in and examine the phase transition process in the view of mutual information, it can describe the phase transition accurately in high temperature limit. However, at low temperature limit this interpretation gets some deviations. Then we take the new saddle into account and introduce a generalized mutual information and find it has a better interpretation of the phase transition process.

    However, this new saddle has a puzzling property that part of its geodesic has a negative length. This is because the endpoints of this geodesic is locating inside the horizon of the bulk. This may lead to the result that the entropy of this saddle is not bounded from below, as we could add arbitrary number
     of the geodesics that has same endpoints. One way to solve ths issue is only one of these coincide geodesics can contribute to the entropy. There might be other ways to solve the problem and that may  worth further researching.

\textbf{Acknowledgment}
This work is supported by NSFC (Grant No.12075246), National Key Research and Development Program of
China (Grant No. 2021YFC2203004), and
the Fundamental Research Funds for the Central Universities. CP is supported by NSFC (Grant No.12175237 and No.12447108 and partly 12247103)

\appendix

\bibliography{references}

\end{document}